\documentclass[pra,aps,showpacs,twocolumn]{revtex4-2}
\usepackage{amsmath}
\usepackage{amssymb}
\usepackage{graphicx}
\usepackage{epstopdf}
\usepackage{mathtools}
\usepackage[bookmarks=false]{hyperref}
\hypersetup{colorlinks=true,citecolor=blue,linkcolor=blue,urlcolor=blue,pdfstartview=FitH,bookmarksopen=true}
\usepackage{xcolor}
\usepackage{soul}
\usepackage{times,txfonts}
\setstcolor{blue}
\setulcolor{red}

\usepackage{physics}

\newcommand{\Rmnum}[1]{\expandafter\@slowromancap\romannumeral #1@}

\begin{document}
	
	\title{Decoherence mitigation for geometric quantum computation}
	\author{X. Y. Sun}
	\affiliation{Department of Physics, Shandong University, Jinan 250100, China}
	\author{P. Z. Zhao}
	\email{pzzhao@sdu.edu.cn}
	\affiliation{Department of Physics, Shandong University, Jinan 250100, China}
	\date{\today}
	
	\begin{abstract}
		Geometric phases depend only on the evolution path determined by the closed circuit in the projective Hilbert space but not on evolution details of the quantum system, leading to geometric quantum computation possessing some intrinsic robustness against control errors. Coordinated with dynamical decoupling, geometric quantum computation admits  additional resilience to the environment-induced decoherence. However, the previous schemes of geometric quantum computation protected by dynamical decoupling require multiple physical qubits to encode a logical qubit, which undoubtedly increases the consumption of physical-qubit resources and the difficulty in the implementation of the  logical-qubit manipulation
		based on physical-qubit driving. In this work, we put forward a scheme of decoherence-mitigated geometric quantum computation based only on physical qubits rather than logical qubits, hence avoiding the additional overhead of physical-qubit resources for logical-qubit encoding as well as the difficulty in the manipulation of logical qubits. Moreover, our scheme focuses on the most general interaction between an individual qubit and its environment so that it mitigates not just dephasing noise but rather regular decoherence. Our proposal thus represents a more realistic and effective approach towards the realization of geometric control with decoherence mitigation. 
	\end{abstract}
	
	\maketitle
	
	\section{Introduction}
	
	In 1984, Berry found that a quantum system initially residing in a nondegenerate eigenstate of the Hamiltonian and then evolving adiabatically and cyclically along a circuit in the parameter space will acquire a geometric phase in addition to the familiar dynamical phase \cite{Berry1984}. In the degenerate case, the acquired geometric phase for the adiabatic and cyclic evolution in the parameter space yields a holonomic matrix, termed as adiabatic non-Abelian geometric phases \cite{Wilczek.Zee1984}. The further extension of geometric phases is from adiabatic evolution  \cite{Tong.Singh.ea2005,Tong2010} to nonadiabatic evolution, where a quantum state undergoing a cyclic evolution will acquire a nonadiabatic geometric phase \cite{Aharonov.Anandan1987} while the cyclic evolution of the state subspace corresponds to a nonadiabatic non-Abelian geometric phase \cite{Anandan1988}. 
	Nonadiabatic geometric phases are dependent only on the evolution path determined by the closed circuit in the projective Hilbert space but independent of evolution details of the quantum system. Due to this geometric robustness, nonadiabatic geometric phases are applied to quantum computation \cite{Wang.Matsumoto2001,Zhu.Wang2002,Sjoqvist.Tong.ea2012,Xu.Zhang.ea2012}, in which the Abelian-phase-based approach is particularly referred to  as geometric quantum computation. It follows that the resulting geometric gates possess some intrinsic robustness against control errors \cite{DeChiara.Palma2003,Carollo.Fuentes-Guridi.ea2004,Solinas.Zanardi.ea2004,Zhu.Wang.ea2005,Lupo.Aniello.ea2007,Filipp.Klepp.ea2009,Thomas.Lababidi.ea2011,Johansson.Sjoqvist.ea2012}. Moreover, this kind of geometric gate avoids the requirement of adiabatic evolution and hence  allows for high-speed implementation. Up to now, many schemes of geometric quantum computation have been proposed \cite{Zhu.Wang2003,Friedenauer.Sjoqvist2003,Solinas.Zanardi.ea2003,Zheng2004,Zhang.Zhu.ea2005,Chen.Feng.ea2006,Cen.Wang.ea2006,Feng.Wang.ea2007,Wu.Wang.ea2007,Kim.Roos.ea2008,Feng.Wu.ea2009,Ota.Goto.ea2009,Thomas.Lababidi.ea2011,Chen.Feng.ea2012,Zhao.Xu.ea2016,Liang.Yue.ea2016,AzimiMousolou2017,Zhao.Cui.ea2017,Chen.Xue2018,Liu.Song.ea2019,Kang.Shi.ea2020,Zhang.Chen.ea2020,Li.Zhao.ea2020,Guo.Yan.ea2020} 
	along with experimental demonstrations on various platforms \cite{Leibfried.DeMarco.ea2003,Du.Zou.ea2006,Kleissler.Lazariev.ea2018,Xu.Hua.ea2020,Zhao.Dong.ea2021a}.   
	
	The robustness of geometric gates lies in the intrinsic geometric resilience to control errors. Accounting for the interaction between the quantum system and its environment, it is necessary to integrate the decoherence-mitigated methods into geometric quantum computation so as to make the quantum gates robust against both control errors and environment-induced decoherence \cite{Cen.Wang.ea2006,Feng.Wu.ea2009,Zhao.Xu.ea2016,Mousolou2018,Chen.Xue.ea2022,Xu.Long2014,Sun.Wang.ea2016,Wu.Zhao2020}.
	The protection of geometric gates with dynamical decoupling represents one of such protocols \cite{Xu.Long2014,Sun.Wang.ea2016,Wu.Zhao2020}. Dynamical decoupling operates by applying external fields to the quantum system, causing some piecewise interactions between the quantum system and its environment to reverse their signs \cite{Viola.Lloyd1998,Viola.Knill.ea1999}. As a consequence, the interaction terms in different durations are compensated by each other and hence canceled out finally \cite{Mukhtar.Saw.ea2010,Chaudhry.Gong2012,Chaudhry.Gong2012a}. Nevertheless, when utilizing dynamical decoupling to protect quantum gates, the decoupling operators not only average out the system-environment interaction but also interfere with the native time evolution of the quantum system. This leads to the time evolution governed by the driving Hamiltonian deviating from the desired trajectory. 
	To realize non-interference protection of quantum gates, one route is to choose certain driving Hamiltonian that is commutable with decoupling operators \cite{Byrd.Lidar2002,Lidar2008,West.Lidar.ea2010,Wu.Sun.ea2023}. As such, the quantum system shall evolve along the desired trajectory determined by the driving Hamiltonian without any interference, meanwhile the system-environment interaction can be canceled through the decoupling operations. Furthermore, if additional confinements are imposed on the time evolution to make the quantum system fulfill the geometric evolution requirement, such as the parallel transport  condition combined with the cyclic evolution condition, the resulting quantum gates are dynamical-decoupling-protected geometric gates. This is just the basic idea of the previous schemes of geometric quantum computation protected by dynamical decoupling.
	
	However, it is highly nontrivial to find a quantum system not only driven by the Hamiltonian that is commutable with the decoupling operators but also fulfilling the geometric evolution requirement. To overcome this challenge, the previous schemes exploited logical qubits to encode quantum information and further chose certain Hamiltonian to realize logical-qubit manipulation \cite{Xu.Long2014,Sun.Wang.ea2016,Wu.Zhao2020}. 
	Because the quantum system needs to satisfy the commutation relation 
	and should be simultaneously complied with the  geometric evolution requirement, the driving Hamiltonian has to keep particular dynamical symmetries and specific forms of qubit-qubit couplings. 
	All these factors undoubtedly increase the consumption of physical-qubit resources and the difficulty of the implementation of logical-qubit manipulation based on restrictedly achievable physical-qubit driving.
	
	In this paper, we present a scheme of geometric quantum computation protected by dynamical decoupling. Our scheme utilizes continuous-wave driving fields so that we realize the decoupling protection of universal geometric gates based on only  physical qubits rather than logical qubits. This avoids the additional resource overhead required for using multiple physical qubits to encode a logical qubit. 
	Considering that the manipulation of physical qubits is much easier than that of logical qubits, our scheme naturally relaxes the restrictions imposed on the driving Hamiltonian for satisfying both the commutation relation and the geometric evolution requirement. Moreover, our scheme focuses on the most general interaction between an individual qubit and its environment, thereby mitigating not just dephasing noise but rather regular decoherence. Our proposal thus  represents a more realistic and effective approach towards the realization of geometric control with decoherence mitigation. 
	
	\section{The outline for dynamical decoupling protection of time evolution}
	
	Before proceeding to the construction of geometric gates, we first outline our basic idea for the protection of time evolution using dynamical decoupling. Consider a quantum system coupled to its environment with the Hamiltonian $	H(t)=H_{0}(t)+H_{\mathrm{E}}+H_{\mathrm{I}}$,
	where $H_0(t)$ denotes the system Hamiltonian, $H_{\mathrm{E}}$ represents the surrounding environment, and $H_{\mathrm{I}}$ is the interaction between the quantum system and its environment. For an individual qubit interacting with its bath, the interaction Hamiltonian is taken as the most general form,
	\begin{align}\label{eq0}
		H_{\mathrm{I}}=\sigma_{x}\otimes{B}_{x}+\sigma_{y}\otimes{B}_{y}+
		\sigma_{z}\otimes{B}_{z},
	\end{align}
	where the left term of the tensor product represents the quantum system and the right term denotes the environment. To eliminate the effect induced by the interaction, we apply an external driving field $H_{\mathrm{c}}(t)$ to the quantum system; hence, the total Hamiltonian is given by $	H_{\mathrm{tot}}(t)=H_{0}(t)+H_{\mathrm{c}}(t)+H_{\mathrm{E}}+H_{\mathrm{I}}$. In the rotating framework with respect to external field $H_{\mathrm{c}}(t)$,
	the total Hamiltonian yields 
	\begin{align}\label{eq00}
		H^{\mathrm{eff}}_{\mathrm{tot}}(t)=U^{\dagger}_{\mathrm{c}}(t)H_{0}(t)U_{\mathrm{c}}(t)+U^{\dagger}_{\mathrm{c}}(t)H_{\mathrm{I}}U_{\mathrm{c}}(t)+H_{\mathrm{E}}.
	\end{align}	
	The evolution operator with respect to  $H_{\mathrm{tot}}(t)$ can be then rewritten as
	\begin{align}\label{eq1}
		U_{\mathrm{tot}}(t)=U_{\mathrm{c}}(t) U^{\mathrm{eff}}_{\mathrm{tot}}(t).
	\end{align} 
	Here, $U_{\mathrm{c}}(t)=\mathcal{T}\exp\bigl[-i\int_{0}^{t}H_{\mathrm{c}}(t^{\prime})\dd{t^\prime}\bigl]$ denotes the rotating operator, $U^{\mathrm{eff}}_{\mathrm{tot}}(t)=\mathcal{T}\exp\bigl[-i\int_{0}^{t}H^{\mathrm{eff}}_{\mathrm{tot}}(t^{\prime})\dd{t^{\prime}}\bigl]$ denotes the unitary operator corresponding to $H^{\mathrm{eff}}_{\mathrm{tot}}(t)$, and $\mathcal{T}$ represents the time ordering. Up to the first order in time, $U^{\mathrm{eff}}_{\mathrm{tot}}(t)$ can be recast as 
	\begin{align}\label{eq2}
		U^{\mathrm{eff}}_{\mathrm{tot}}(t)=&[\mathcal{T}e^{-i\int^{t}_{0}U^{\dagger}_{\mathrm{c}}(t^{\prime})H_{0}(t^{\prime})U_{\mathrm{c}}(t^{\prime})\dd{t^{\prime}}}\otimes{U}_{\mathrm{E}}(t)]\notag\\
		&\times e^{-i\int^{t}_{0}U^{\dagger}_{\mathrm{c}}(t^{\prime})H_{\mathrm{I}}U_{\mathrm{c}}(t^{\prime})\dd{t^{\prime}}}+O(t^{2}),
	\end{align}
	where ${U}_{\mathrm{E}}(t)=\exp(-iH_{\mathrm{E}}t)$ denotes the time evolution induced purely by the bath Hamiltonian. 
	
	If the external field is engineered to satisfy the following two conditions,
	\begin{align}\label{eq3}
		&(\mathrm{a})~~U_{\mathrm{c}}(t+\tau)=U_{\mathrm{c}}(t),
		\notag\\
		&(\mathrm{b})~~\int^{\tau}_{0}U^{\dagger}_{\mathrm{c}}(t)H_{\mathrm{I}}U_{\mathrm{c}}(t)\dd{t}=0,
	\end{align}
	we can conclude that the quantum system undergoing a period of time $\tau$ shall be decoupled from its surrounding environment up to the leading order. 
	The condition $(\mathrm{a})$ implies that $U_{\mathrm{c}}(\tau)=U_{\mathrm{c}}(0)=I$, where $I$ represents the identity operator. This indicates that after a period of the time $\tau$, the total evolution operator in Eq.~(\ref{eq1}) is reduced to $U_{\mathrm{tot}}(\tau)= U^{\mathrm{eff}}_{\mathrm{tot}}(\tau)$. Furthermore, the condition $(\mathrm{b})$ ensures that the effect of the system-environment interaction is eliminated from $U^{\mathrm{eff}}_{\mathrm{tot}}(\tau)$ up to the first order, seen from Eq.~(\ref{eq2}). 
	As a consequence, the evolution operator of the quantum system is reduced to
	\begin{align}\label{eq}
		\mathcal{U}(\tau)=\mathcal{T}e^{-i\int^{\tau}_{0}U^{\dagger}_{\mathrm{c}}(t)H_{0}(t)U_{\mathrm{c}}(t)\dd{t}}.
	\end{align} 
	If we further adjust the driving Hamiltonian $H_{0}(t)$ to make the effective Hamiltonian $H^{\mathrm{eff}}_{\mathrm{S}}(t)\equiv U^{\dagger}_{\mathrm{c}}(t)H_{0}(t)U_{\mathrm{c}}(t)$ generate a geometric evolution, the resulting unitary  operator $\mathcal{U}(\tau)$ yields a decoherence-mitigated geometric gate. In the following, we demonstrate how to realize a universal set of geometric gates, including arbitrary one-qubit gates and a nontrivial two-qubit gate, through engineering the external field cooperating with the adjustment of the driving Hamiltonian of the quantum system. 
	
	\section{One-qubit gates}
	
	As outlined above, to realize the dynamical decoupling protection of geometric gates, we first apply an external driving field $H_{\mathrm{c}}(t)$ to the native time evolution governed by the driving Hamiltonian $H_{0}(t)$ and then process the dynamics of the quantum system under the framework with respect to $H_{\mathrm{c}}(t)$. From Eqs.~(\ref{eq00}) and (\ref{eq1}), we conclude that if the unitary operator $U_{\mathrm{c}}(t)$ corresponding to  $H_{\mathrm{c}}(t)$ satisfies the conditions $(\mathrm{a})$ and $(\mathrm{b})$ in Eq.~(\ref{eq3}), the total evolution operator yields the one in Eq.~(\ref{eq}) determined by the effective Hamiltonian $H^{\mathrm{eff}}_{\mathrm{S}}(t)$ along with the elimination of the leading-order effect from the interaction Hamiltonian.
		Therefore, the key point for the realization of the dynamical decoupling protection of geometric gates is to engineer the unitary operator $U_{\mathrm{c}}(t)$ that satisfies conditions $(\mathrm{a})$ and $(\mathrm{b})$ and to design the effective Hamiltonian $H^{\mathrm{eff}}_{\mathrm{S}}(t)$ for implementing the geometric evolution. 
		This being done, the real driving Hamiltonian $H_{0}$ combined with the compatible external driving field $H_{\mathrm{c}}(t)$ can be then obtained according to the relations $H_{0}(t)=U_{\mathrm{c}}(t)H_{\mathrm{S}}^{\mathrm{eff}}(t)U^{\dagger}_{\mathrm{c}}(t)$ 
		and   
		$H_{\mathrm{c}}(t)=i\dot{U}_{\mathrm{c}}(t)U^{\dagger}_{\mathrm{c}}(t)$.
		In the following, we first demonstrate how to  engineer the unitary operator $U_{\mathrm{c}}(t)$ that satisfies the conditions $(\mathrm{a})$ and $(\mathrm{b})$, and then we present how to design the effective Hamiltonian $H^{\mathrm{eff}}_{\mathrm{S}}(t)$ to realize one-qubit geometric gates.
	
	To satisfy the condition $(\mathrm{a})$, a natural choice of the unitary operator is the periodical function $U^{\mu}_{\mathrm{c}}(t)=\exp\bigl(-i2\pi{n}\sigma_{\mu}t/\tau\bigl)$, with the positive integer $n$ and the Pauli operator $\sigma_{\mu(=x,y,z)}$. Clearly, this function starting from $t=t_0$ shall return to itself after a period of time evolution $t=t_{0}+\tau$. To further satisfy the condition $(\mathrm{b})$, we first consider the longitudinal noise, i.e., the terms related to $\sigma_{x}$ and $\sigma_{y}$ in Eq.~(\ref{eq0}). In this case, we can take $U^{\mu}_{\mathrm{c}}(t)$ to be $U^{z}_{\mathrm{c}}(t)=\exp\bigl(-i2\pi{n}\sigma_{z}t/\tau\bigl)$. It should operate because
	$U^{z\dagger}_{\mathrm{c}}(t)\sigma_{x}U^{z}_{\mathrm{c}}(t)=\cos(4\pi nt/\tau)\sigma_{x}-\sin(4\pi nt/\tau)\sigma_{y}$, $U^{z\dagger}_{\mathrm{c}}(t)\sigma_{y}U^{z}_{\mathrm{c}}(t)=\sin(4\pi nt/\tau)\sigma_{x}+\cos(4\pi nt/\tau)\sigma_{y}$,
	and the integral of coefficients over a period of time $\tau$ is 0.
	If we further take into account the transverse noise, i.e., the term related to $\sigma_{z}$ in Eq.~(\ref{eq0}), we need to additionally resort to $U^{x}_{\mathrm{c}}(t)$ or $U^{y}_{\mathrm{c}}(t)$. Considering that the integral of the above mentioned coefficients is still 0 when $2\pi$ in the exponential of the unitary operator is replaced by $\pi$, we finally take a trial unitary operator for suppressing all the terms in the interaction Hamiltonian as
	\begin{align}\label{eq4}
		U_{\mathrm{c}}(t)=e^{-i\pi n_x\sigma_xt/\tau}e^{-i\pi n_z\sigma_zt/\tau}, 	
	\end{align}
	where $n_x$ and $n_z$ are simultaneously set as positive odd or even numbers for satisfying the periodical condition with $n_x\neq n_z$.
	It is clear that
	$U_{\mathrm{c}}(t+\tau)=e^{-i\pi n_x\sigma_xt/\tau}e^{-i\pi n_{x}\sigma_{x}}e^{-i\pi n_z\sigma_zt/\tau}e^{-i\pi n_{z}\sigma_{z}}=U_{\mathrm{c}}(t)$, i.e., the condition $(\mathrm{a})$ is fulfilled. 
	Next, we demonstrate in detail that 
	the chosen $U_{\mathrm{c}}(t)$ also satisfies the condition $(\mathrm{b})$
	aiming at a general form of an individual qubit coupled to its environment with the interaction Hamiltonian depicted by Eq.~(\ref{eq3}).
	For this, we substitute Eqs.~(\ref{eq0}) and (\ref{eq4}) into $U^{\dagger}_{\mathrm{c}}(t)H_{\mathrm{I}}U_{\mathrm{c}}(t)$, and then we have
	\begin{align}
		U^{\dagger}_{\mathrm{c}}(t)H_{\mathrm{I}}U_{\mathrm{c}}(t)
		=&U^{\dagger}_{\mathrm{c}}(t)\sigma_{x}U_{\mathrm{c}}(t)\otimes B_{x}+U^{\dagger}_{\mathrm{c}}(t)\sigma_{y}U_{\mathrm{c}}(t)\otimes B_{y}
		\notag\\
		&+U^{\dagger}_{\mathrm{c}}(t)\sigma_{z}U_{\mathrm{c}}(t)\otimes B_{z}
	\end{align}
	along with
	\begin{align}
		U^{\dagger}_{\mathrm{c}}(t)\sigma_{x}U_{\mathrm{c}}(t)=&\cos(2\pi n_zt/\tau)\sigma_{x}-\sin(2\pi n_zt/\tau)\sigma_{y},
		\notag\\
		U^{\dagger}_{\mathrm{c}}(t)\sigma_{y}U_{\mathrm{c}}(t)=&\cos(2\pi n_xt/\tau)\sin(2\pi n_zt/\tau)\sigma_{x}  \notag\\
		&+\cos(2\pi n_xt/\tau)\cos(2\pi n_zt/\tau)\sigma_{y}  \notag\\
		&-\sin(2\pi n_xt/\tau)\sigma_{z},
		\notag\\
		U^{\dagger}_{\mathrm{c}}(t)\sigma_{z}U_{\mathrm{c}}(t)=&\sin(2\pi n_xt/\tau)\sin(2\pi n_zt/\tau)\sigma_{x}   \notag\\
		&+\sin(2\pi n_xt/\tau)\cos(2\pi n_zt/\tau)\sigma_{y}  \notag\\
		&+\cos(2\pi n_xt/\tau)\sigma_{z}.
	\end{align}
	For the terms $\cos(2\pi nt/\tau)$ and $\sin(2\pi nt/\tau)$, with $n=n_x$ or $n_z$, it is easy to verify that $\int^{\tau}_{0}\cos(2\pi nt/\tau)\dd{t}=\int^{\tau}_{0}\sin(2\pi nt/\tau)\dd{t}=0$. 
	Furthermore, accounting for 
	$\cos(2\pi n_xt/\tau)\cos(2\pi n_zt/\tau)=\cos[2\pi (n_x+n_z)t/\tau]/2+\cos[2\pi (n_x-n_z)t/\tau]/2$ and $n_{x}\neq n_{z}$, it is clear that the integral of $\cos(2\pi n_xt/\tau)\cos(2\pi n_zt/\tau)$ over $[0,\tau]$ is equal to 0.
	Similarly, we can conclude that the integrals of the other terms $\sin(2\pi n_xt/\tau)\cos(2\pi n_zt/\tau)$, $\cos(2\pi n_xt/\tau)\sin(2\pi n_zt/\tau)$ and $\sin(2\pi n_xt/\tau)\sin(2\pi n_zt/\tau)$ are all equal to 0. Therefore, we have 
	$\int^{\tau}_{0}U^{\dagger}_{\mathrm{c}}(t)H_{\mathrm{I}}U_{\mathrm{c}}(t)\dd{t}=0$, i.e., the condition $(\mathrm{b})$ is fulfilled. To sum up, both the conditions $(\mathrm{a})$ and $(\mathrm{b})$ are satisfied, and the chosen $U_{\mathrm{c}}(t)$ is a legitimate external field.
	
	Having found an external driving field fulfilling the conditions $(\mathrm{a})$ and $(\mathrm{b})$, let us now construct the effective Hamiltonian $H_{\mathrm{S}}^{\mathrm{eff}}(t)=U^{\dagger}_{\mathrm{c}}(t)H_{0}(t)U_{\mathrm{c}}(t)$ to generate an arbitrary geometric one-qubit gate.
	To this end, we consider a projective Hilbert space spanned  by
	\begin{align}\label{eq5}
		\ket{\phi_1(t)}&=\cos\frac{\theta(t)}{2}\ket{0}+\sin\frac{\theta(t)}{2}e^{i\varphi(t)}\ket{1},\notag\\
		\ket{\phi_2(t)}&=\sin\frac{\theta(t)}{2}e^{-i\varphi(t)}\ket{0}-\cos\frac{\theta(t)}{2}\ket{1},
	\end{align}
	where $\theta(t)$ and $\varphi(t)$ are two time-dependent parameters with the requirement $\theta(0)=\theta(\tau)\equiv\theta_{0}$ and
	$\varphi(0)=\varphi(\tau)\equiv\varphi_{0}$.
	Clearly, $\ket{\phi_k(\tau)}=\ket{\phi_k(0)}$ after an evolution period $\tau$. Arranged with this projective Hilbert space, we can construct another set of orthonormal basis states $\{\ket{\psi_{k}(t)}\}^{2}_{k=1}$ that 
	satisfy the Schr{\"o}dinger equation
	$i|\dot{\psi}_{k}(t)\rangle=H_{\mathrm{S}}^{\mathrm{eff}}(t)\ket{\psi_{k}(t)}$.
	The basis state $\ket{\psi_k(t)}$ is defined by  $\ket{\psi_k(t)}\equiv\exp[i\gamma_k(t)]\ket{\phi_k(t)}$ with the phase chosen as $\gamma_{k}(t)=i\int^{t}_{0}\bra{\phi_{k}(t^{\prime})}\dot{\phi}_{k}(t^{\prime})\rangle\dd{t^{\prime}}$. It is easy to verify that $\ket{\psi_{k}(0)}=\ket{\phi_{k}(0)}$ and $\ket{\psi_{k}(\tau)}=\exp[i\gamma_k(\tau)]\ket{\phi_{k}(0)}$.
	According to the Schr{\"o}dinger equation, we can obtain the form of the Hamiltonian $H_{\mathrm{S}}^{\mathrm{eff}}(t)=i\sum_{k}|{\dot{\psi}_{k}(t)}\rangle\bra{\psi_{k}(t)}$ and hence we have
	\begin{align}\label{eq6}
		H_{\mathrm{S}}^{\mathrm{eff}}(t)=i\sum_{l\neq k}\bra{\phi_l(t)}\dot{\phi}_k(t)\rangle\ket{\phi_l(t)}\bra{\phi_k(t)}.
	\end{align}
	The time evolution governed by the Hamiltonian $H_{\mathrm{S}}^{\mathrm{eff}}(t)$ reads $\mathcal{U}(t)=\sum_{k}\ket{\psi_{k}(t)}\bra{\psi_{k}(0)}$. After an evolution period $\tau$, the unitary operator then yields
	\begin{align}\label{eq7}
		\mathcal{U}(\tau)=e^{-i\gamma(\tau)}\ket{\phi_1(0)}\bra{\phi_1(0)}+e^{i\gamma(\tau)}\ket{\phi_2(0)}\bra{\phi_2(0)}, 
	\end{align}
	where the phase $\gamma(\tau)=i\int_{0}^{\tau}[1-\cos\theta(t)]\dot{\varphi}(t)\dd{t}/2$ is obtained by inserting Eq.~(\ref{eq5}) into the expressions of $\gamma_{1}(\tau)$ and $\gamma_{2}(\tau)$.
	Clearly then, starting from one of the basis states $\ket{\phi_1(0)}$ and $\ket{\phi_2(0)}$, the quantum system governed by the Hamiltonian $H_{\mathrm{S}}^{\mathrm{eff}}(t)$, undergoing a cyclic evolution with period time $\tau$ shall acquire a phase $-\gamma(\tau)$ or $\gamma(\tau)$. 
	Using the expression in Eq.~(\ref{eq6}), it is easy to verify
	\begin{align}
		\bra{\psi_k(t)}H^{\mathrm{eff}}_\mathrm{S}(t)\ket{\psi_k(t)}=\bra{\phi_k(t)}H^{\mathrm{eff}}_\mathrm{S}(t)\ket{\phi_k(t)}=0. 
	\end{align}
	This indicates that the time evolution for  state $\ket{\psi_k(t)}$ is parallel transport with the removal of dynamical phases and hence the phases $\pm\gamma(\tau)$ are purely geometric phases.
	Therefore, the evolution operator described in Eq.~(\ref{eq7}) is a geometric gate.
	By substituting Eq.~(\ref{eq5}) into Eq.~(\ref{eq7}), we can verify that the unitary operator can be rewritten as $\mathcal{U}(\tau)=\exp[-i\gamma(\tau)\boldsymbol{\mathrm{n}\cdot\sigma}]$, where $\mathbf{n}=(\sin\theta_0\cos\varphi_0, \sin\theta_0\sin\varphi_0, \cos\theta_0)$ is an arbitrary unit vector determining the orientation of a rotation axis
	and $\boldsymbol{\sigma}=(\sigma_x, \sigma_y, \sigma_z)$ is the standard Pauli operator.
	Obviously, the unitary operator represents an arbitrary one-qubit gate along an arbitrary rotation axis with an arbitrary rotation angle.
	It is exactly the quantum gate that we aim to realize.
	
	In the above discussions, we have found a cyclical external field $H_{\mathrm{c}}(t)$ determined by Eq.~(\ref{eq4}) to compensate the system-environment interaction described by Eq.~(\ref{eq0}). In this case, the evolution operator after applying the external field is reduced to the unitary operator $\mathcal{U}(\tau)$ in Eq.~(\ref{eq}) at the final time $\tau$. The unitary operator is completely decided by the effective Hamiltonian $H_{\mathrm{S}}^{\mathrm{eff}}(t)=U^{\dagger}_{\mathrm{c}}(t)H_{0}(t)U_{\mathrm{c}}(t)$. Afterwards, we constructed the effective Hamiltonian $H_{\mathrm{S}}^{\mathrm{eff}}(t)$ as the form in Eq.~(\ref{eq6}) along with the projective basis given by Eq.~(\ref{eq5}). As such, the final unitary operator yields a geometric gate expressed as Eq.~(\ref{eq7}). It is worth noting that $H_{\mathrm{S}}^{\mathrm{eff}}(t)$ is the Hamiltonian under the rotating framework but not a real driving field. 
	To realize geometric gates protected by dynamical decoupling, we need to achieve the real driving Hamiltonian $H_{0}(t)$ combined with the external field $H_{\mathrm{c}}(t)$. This can be done through the aforementioned relations
	$H_{0}(t)=U_{\mathrm{c}}(t)H_{\mathrm{S}}^{\mathrm{eff}}(t)U^{\dagger}_{\mathrm{c}}(t)$ and $H_{\mathrm{c}}(t)=i\dot{U}_{\mathrm{c}}(t)U^{\dagger}_{\mathrm{c}}(t)$ along with $U_{\mathrm{c}}(t)$ given by Eq.~(\ref{eq4}) and $H_{\mathrm{S}}^{\mathrm{eff}}(t)$ obtained from Eqs.~(\ref{eq5}) and (\ref{eq6}).
	As a consequence, the total driving field $H_{\mathrm{S}}(t)\equiv{H}_{0}(t)+H_{\mathrm{c}}(t)$ arrives at 
	\begin{align}
		H_\mathrm{S}(t)=&\Omega_x(t)\sigma_x+\Omega_y(t)\sigma_y+\Omega_z(t)\sigma_z,
	\end{align}
	with the parameters $\Omega_x(t)$, $\Omega_y(t)$, and $\Omega_z(t)$ given by
	\begin{widetext}
	\begin{align}\label{eq9}
		\Omega_x(t)=&-\frac{\dot{\theta}(t)}{2}\sin[\varphi(t)+2n_z\pi t/\tau]
		-\frac{\dot{\varphi}(t)}{4}\sin2\theta(t)\cos[\varphi(t)+2n_z\pi t/\tau]+\pi n_x/\tau, \notag\\
		\Omega_y(t)=&\frac{\dot{\theta}(t)}{2}\cos(2n_x\pi t/\tau)\cos[\varphi(t)+2n_z\pi t/\tau] 
		-\frac{\dot{\varphi}(t)}{4}\{\sin2\theta(t)\cos(2n_x\pi t/\tau)\sin[\varphi(t)+2n_z\pi t/\tau]+[1-\cos2\theta(t)]\notag\\
		&\times\sin(2n_x\pi t/\tau)\}
		-n_z\pi\sin(2n_x\pi t/\tau)/\tau, \notag\\
		\Omega_z(t)=&\frac{\dot{\theta}(t)}{2}\sin(2n_x\pi t/\tau)\cos[\varphi(t)+2n_z\pi t/\tau] -\frac{\dot{\varphi}(t)}{4}
		\{\sin2\theta(t)\sin(2n_x\pi t/\tau)\sin[\varphi(t)+2n_z\pi t/\tau]
		-[1-\cos2\theta(t)]  \notag\\
		&\times\cos(2n_x\pi t/\tau)\}+n_z\pi\cos(2n_x\pi t/\tau)/\tau.
	\end{align}
	\end{widetext}
	This Hamiltonian describes a two-level system driven by an off-resonant laser with the detuning $\Delta(t)\equiv2\Omega_{z}(t)$ and the complex Rabi frequency $\Omega_{0}(t)\equiv2[\Omega_{x}(t)+i\Omega_{y}(t)]$. It can be realized in many physical systems, such as trapped ions \cite{Cirac.Zoller1995}, superconducting circuits \cite{Blais.Gambetta.ea2007,Koch.Yu.ea2007}, and Rydberg atoms \cite{Jaksch.Cirac.ea2000,Lukin.Fleischhauer.ea2001,Saffman.Walker.ea2010}.
	
	In the practical realization, we need to first confirm the parameters $\theta(t)$ and $\varphi(t)$ along with $n_{x}$, $n_{z}$, and the evolution period $\tau$ according to the desired quantum gate. Based on the given parameters, we can then obtain the physical parameters $\Omega_x(t)$, $\Omega_y(t)$, and $\Omega_z(t)$ by following the expressions  in Eq.~(\ref{eq9}). In such a way, we finally get the total driving Hamiltonian $H_{\mathrm{S}}(t)$ for the realization of geometric gates protected by dynamical decoupling. Let us take an example, $\mathcal{U}(\tau)=\exp(-i\pi \sigma_{z}/8)$ with $\gamma(\tau)=\pi/8$, to specifically illustrate the realization of our geometric gate. To this end, we choose the evolution path traced by  $[\theta(t),\varphi(t)]$ starting from $\theta(0)=0$ to $\theta(\tau/2)=\pi$, with $\varphi(t)=0$, and then returning from $\theta(\tau/2)=\pi$ to $\theta(\tau)=0$, with $\varphi(t)=\pi/8$.
	Here, $\theta(t)$ can be an arbitrary function confined by the above designing. For simplicity, we can take $\theta(t)$ to be a piecewise function such that $\theta(t)=2\pi{t}/\tau$ in the interval $t\in[0,\tau/2)$ and $\theta(t)=2\pi(\tau-t)/\tau$ in the interval $t\in[\tau/2,\tau]$.
	Additionally, we take $n_{x}=1$ and $n_{z}=3$.
	In this case, the parameters of the total driving field during $[0,\tau/2)$ are obtained as  
	$\Omega_x(t)=-\pi\sin(6\pi t/\tau)/\tau+\pi/\tau$, $\Omega_y(t)=\pi\cos(2\pi t/\tau)\cos(6\pi t/\tau)/\tau-3\pi\sin(2\pi t/\tau)/\tau$, and
	$\Omega_z(t)=\pi\sin(2\pi t/\tau)\cos(6\pi t/\tau)/\tau+3\pi\cos(2\pi t/\tau)/\tau$,  
	and the parameters during $[\tau/2,\tau]$ are obtained as
	$\Omega_x(t)=\pi\sin(\pi/8+6\pi t/\tau)/\tau+\pi/\tau$, $\Omega_y(t)=-\pi\cos(2\pi t/\tau)\cos(\pi/8+6\pi t/\tau)/\tau-3\pi\sin(2\pi t/\tau)/\tau$, and
	$\Omega_z(t)=-\pi\sin(2\pi t/\tau)\cos(\pi/8+6\pi t/\tau)/\tau+3\pi\cos(2\pi t/\tau)/\tau$.
	Driven by such a Hamiltonian for a period of time $\tau$, the desired gate $\mathcal{U}(\tau)=\exp(-i\pi \sigma_{z}/8)$ can be realized. 
	
	It is worth noting that our decoupling scheme needs to consume some additional resources. To visualize this point, we compare our scheme with the one excluding decoupling protection by taking $\mathcal{U}(\tau)=\exp(-i\pi \sigma_{z}/8)$ as an example. The scheme without protection corresponds to  $n_{x}=n_{z}=0$ and hence the driving Hamiltonian yields $H_{\mathrm{S}}=-\pi\sigma_{y}/\tau$ during $[0,\tau/2)$ and $H_{\mathrm{S}}=\pi\sin(\pi/8)\sigma_{x}/\tau-\pi\cos(\pi/8)\sigma_{y}/\tau$ during $[\tau/2,\tau]$. The pulse envelope $\sqrt{|\Omega_{x}|^2+|\Omega_{y}|^2+|\Omega_{z}|^2}$ that describes the absolute values of an eigenenergy is then given by $\Omega\equiv\pi/\tau$. Compared with such a scheme, our scheme needs a larger pulse envelope. For illustrating this, we take our driving Hamiltonian in the interval $[0,\tau/2)$ as an example and rewrite it as $H_{\mathrm{S}}(t)=H_{\mathrm{S_{1}}}(t)+H_{\mathrm{S_{2}}}(t)+H_{\mathrm{S_{3}}}(t)$,  with $H_{\mathrm{S_{1}}}(t)\equiv-\pi\sin(6\pi{t}/\tau)\sigma_{x}/\tau+\pi\cos(2\pi{t}/\tau)\cos(6\pi{t}/\tau)\sigma_{y}/\tau+\pi\sin(2\pi{t}/\tau)\cos(6\pi{t}/\tau)\sigma_{z}/\tau$, $H_{\mathrm{S_{2}}}(t)\equiv\pi\sigma_{x}/\tau-\pi\sin(2\pi{t}/\tau)\sigma_{y}/\tau+\pi\cos(2\pi{t}/\tau)\sigma_{z}/\tau$, and $H_{\mathrm{S_{3}}}(t)\equiv-2\pi\sin(2\pi{t}/\tau)\sigma_{y}/\tau+2\pi\cos(2\pi{t}/\tau)\sigma_{z}/\tau$. The pulse envelopes corresponding to $H_{\mathrm{S_{1}}}(t)$,  $H_{\mathrm{S_{2}}}(t)$, and  $H_{\mathrm{S_{3}}}(t)$ are then given by $\Omega$, $\sqrt{2}\Omega$, and $2\Omega$.
		Clearly, our scheme needs a larger pulse envelope. This indicates that for a fixed  evolution period $\tau$,  our scheme generally requires an increased pulse amplitude. Additionally, our scheme needs to introduce extra fields, such as  $H_{\mathrm{S_{2}}}(t)$ and $H_{\mathrm{S_{3}}}(t)$.
			
\begin{figure}[t]
			\includegraphics[scale=0.58]{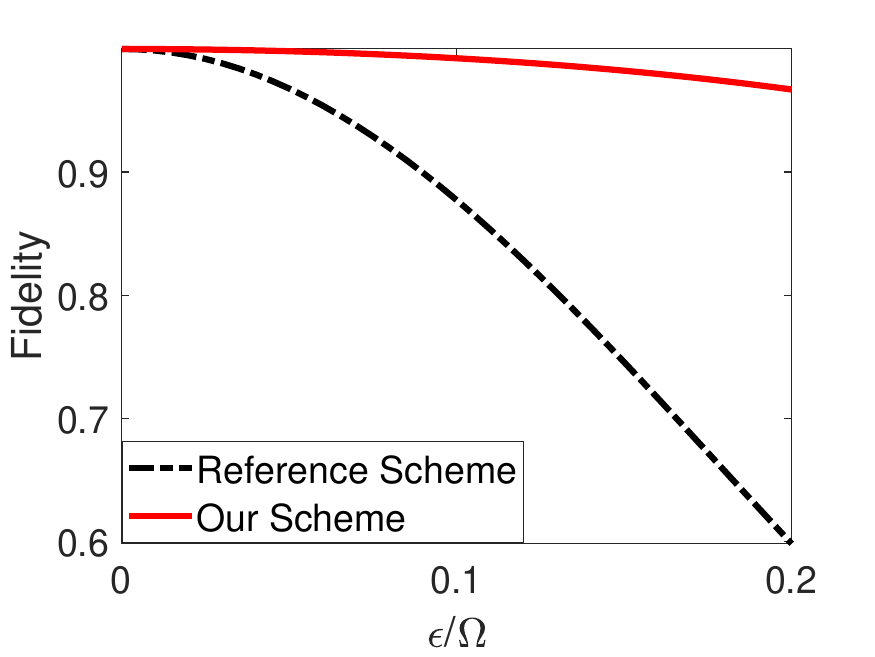}
			\caption{The fidelity of one-qubit gate $\exp(-i\pi \sigma_{z}/8)$ for our scheme (red line) and the reference scheme without decoupling protection (black line) versus the noise strength $\epsilon$ over $[0,0.2\Omega]$.}
			\label{Fig1}
		\end{figure}

	To demonstrate the enhancement of our scheme, we numerically compare the performance of our approach with the  
		one excluding decoupling protection. The performance is characterized by the fidelity $F\equiv\bra{\psi}\rho\ket{\psi}$, where $\ket{\psi}$ is the target state and $\rho$ is the real state under noises. In our numerical simulations, we still take the  gate $\mathcal{U}(\tau)=\exp(-i\pi \sigma_{z}/8)$. The total driving Hamiltonian for the implementation of such a gate along with the parameter design is engineered by following the aforementioned method. The interaction Hamiltonian is taken as the Heisenberg coupling
		$H_{I}=\epsilon(\sigma^{s}_{x}\sigma^{e}_{x}+\sigma^{s}_{y}\sigma^{e}_{y}+\sigma^{s}_{z}\sigma^{e}_{z})$ with the noise strength $\epsilon$, where the left term of the tensor product acts on the system qubit and the right term of the tensor product acts on the environment qubit.	
		Furthermore, we take the initial state as $(\ket{0}+\ket{1})/\sqrt{2}$ and the pulse parameter $\Omega=2\pi \ \time100\mathrm{MHz}$.
		In Fig.~\ref{Fig1}, we present the fidelity $F$ versus the noise strength $\epsilon\in[0,0.2\Omega]$ for our scheme and the reference scheme without decoupling protection, represented by the red and black lines. It shows that our scheme notably enhances the fidelity of quantum gates. For example, in the case $\epsilon=0.2\Omega$ where the fidelity of the bare gate is lower than $60\%$, our scheme improves the gate fidelity to $96.70\%$.

	\section{Two-qubit gates}
	
	In the above section, we have realized an arbitrary geometric one-qubit gate protected by dynamical decoupling. To realize universal quantum computation, we also need a nontrivial two-qubit gate. In the following, we demonstrate how to realize a nontrivial geometric two-qubit gate protected by dynamical decoupling. For this, we consider the following driving Hamiltonian with the qubit-qubit interaction 
	\begin{equation}\label{eq10}
		H_{0}(t)=J_1(t)(\sigma_x\sigma_x+\sigma_y\sigma_y)+J_2(t)(\sigma_x\sigma_y-\sigma_y\sigma_x), 
	\end{equation}
	where the first term represents the $XY$ interaction and the second term represents the Dzialoshinski-Moriya interaction with the coupling parameters $J_1(t)$ and $J_2(t)$.
    This Hamiltonian can be realized using trapped ions with the S{\o}rensen-M{\o}lmer setting \cite{Sorensen.Molmer1999a,Sorensen.Molmer2000}.
		Specifically, we use a pair of blue sideband lasers with the same detuning $-(\nu+\delta)$ and different Rabi frequencies $\Omega_{1}(t)$ and $\Omega_{2}(t)$ to respectively drive the two ions, where $\nu$ is the vibrational frequency and $\delta$ is an additional detuning. In the rotating frame and the rotating-wave approximation, the Hamiltonian of this quantum system reads $H(t)=i(\eta/2)\exp(-i\delta t)[\Omega_{1}(t)a^{\dagger}\ket{1}_{1}\bra{0}	+\Omega_{2}(t)a^{\dagger}\ket{1}_{2}\bra{0}]+\mathrm{H.c.}$ in the Lamb-Dicke regime. Here, $a$ and $a^{\dagger}$ are the annihilation and creation operators of the vibrational mode and $\eta$ is the Lamb-Dicke parameter. If the large detuning approximation  $\delta\gg\eta\Omega_{1}(t)/2,\eta\Omega_{2}(t)/2$ is satisfied, the Hamiltonian yields an effective one: $H(t)=\Omega_{\mathrm{eff}}(t)\ket{01}\bra{10}+\mathrm{H.c.}$, with $\Omega_{\mathrm{eff}}(t)=\eta^{2}\Omega^{\ast}_{1}(t)\Omega_{2}(t)/(4\delta)$. Here, we have ignored the Stark shift terms that can be easily compensated by using additional lasers. The effective Hamiltonian describes the $XY$ interaction if $\Omega_{\mathrm{eff}}(t)$ is a real number, and it describes the Dzialoshinski-Moriya interaction if  $\Omega_{\mathrm{eff}}(t)$ is an imaginary number. Therefore, the interactions described above are realistically available in trapped ions.
	
	To suppress the system-environment interaction described in Eq.~(\ref{eq0}), we resort to the periodic decoupling sequence $\{\sigma_{0}^{\otimes2},\sigma_{1}^{\otimes2},\sigma_{2}^{\otimes2},\sigma_{3}^{\otimes2}\}$ with the definitions $\sigma_{0}\equiv{I}$, $\sigma_{1}\equiv\sigma_{x}$, $\sigma_{2}\equiv\sigma_{y}$, and $\sigma_{3}\equiv\sigma_{z}$.
	This is because the interaction Hamiltonian under the action of the periodic decoupling sequence will be averaged out such that  
	$\sum^{3}_{\alpha=0}\sigma_{\alpha}H_{\mathrm{I}}\sigma_{\alpha}=\sigma_{x}\otimes{B}_{x}+\sigma_{y}\otimes{B}_{y}+\sigma_{z}\otimes{B}_{z}+\sigma_{x}\otimes{B}_{x}-\sigma_{y}\otimes{B}_{y}-
	\sigma_{z}\otimes{B}_{z}-\sigma_{x}\otimes{B}_{x}+\sigma_{y}\otimes{B}_{y}-
	\sigma_{z}\otimes{B}_{z}-\sigma_{x}\otimes{B}_{x}-\sigma_{y}\otimes{B}_{y}+
	\sigma_{z}\otimes{B}_{z}=0$.
	If we insert the periodic decoupling sequence into the time evolution of the quantum system, the unitary operator over the evolution time  $4\tau$ reads
	\begin{align}
		U_{\mathrm{tot}}(4\tau)=&\prod_{k=0}^3\sigma_k^{\otimes2}\mathcal{T}e^{-i\int_{k\tau}^{(k+1)\tau}H(t)\dd{t}}\sigma_k^{\otimes2} \notag\\
		=&\prod_{k=0}^3\mathcal{T}e^{-i\int_{k\tau}^{(k+1)\tau}\sigma_k^{\otimes2}H_{0}(t)\sigma_k^{\otimes2}\dd{t}}\otimes{e^{-i4H_{\mathrm{E}}\tau}}
		\notag\\
		&\times e^{-i\sum^{3}_{\alpha=0}\sigma_\alpha^{\otimes2}H_{\mathrm{I}}\sigma^{\otimes2}_\alpha\tau}+O(\tau^2).
	\end{align}
	Note that $\sum^{3}_{\alpha=0}\sigma_\alpha^{\otimes2}H_{\mathrm{I}}\sigma^{\otimes2}_\alpha=0$ as mentioned above and hence the unitary operator yields
	\begin{align}
		U_{\mathrm{tot}}(4\tau)	=\prod_{k=0}^3\mathcal{T}e^{-i\int_{k\tau}^{(k+1)\tau}H^{\mathrm{eff}}_{\mathrm{S}}(t)\dd{t}}\otimes{e^{-i4H_{\mathrm{E}}\tau}}
		+O(\tau^2).
	\end{align}
	Here, $H^{\mathrm{eff}}_{\mathrm{S}}(t)$ represents the effective Hamiltonian which is a piecewise function defined as $H^{\mathrm{eff}}_{\mathrm{S}}(t)\equiv\sigma_k^{\otimes2}H_{0}(t)\sigma_k^{\otimes2}$.
	The above equation indicates that after applying the periodic decoupling sequence, the quantum system is completely decoupled from its environment up to the first order but the driving Hamiltonian is altered by the decoupling operators to a piecewise effective Hamiltonian. The resulting unitary operator is then recast as
	\begin{align}
		\mathcal{U}(4\tau)=\prod_{k=0}^3\mathcal{T}e^{-i\int_{k\tau}^{(k+1)\tau}H^{\mathrm{eff}}_{\mathrm{S}}(t)\dd{t}}
	\end{align}   
	
	Let us now construct the effective Hamiltonian 
	$H^{\mathrm{eff}}_{\mathrm{S}}(t)$ to generate a geometric two-qubit gate. For this, we divide the whole evolution time into three intervals. In the first and third intervals $t\in[0,\tau)\cup(3\tau,4\tau]$, we take the parameters $J_{1}(t)=J(t)/2$ and $J_{2}(t)=0$.
	In the second interval $t\in[\tau,3\tau]$, we take $J_{1}(t)=J(t)\cos(\pi-\gamma)/2$ and $J_{2}(t)=J(t)\sin(\pi-\gamma)/2$.
	In this case, the piecewise Hamiltonian reads 
	\begin{align}
		H^{\mathrm{eff}}_{\mathrm{S}}(t)=J(t)R_{x}
	\end{align}
	for $t\in[0,\tau)\cup(3\tau,4\tau]$ and
	\begin{align}
		H^{\mathrm{eff}}_{\mathrm{S}}(t)=J(t)[\cos(\pi-\gamma)R_{x}+\sin(\pi-\gamma)R_{y}]
	\end{align}
	for $t\in[\tau,3\tau]$, where $R_{x}\equiv\ket{01}\bra{10}+\ket{10}\bra{01}$
	and $R_{y}\equiv-i\ket{01}\bra{10}+i\ket{10}\bra{01}$. 
	It is clear that in the computational space, the basis states $\ket{01}$ and $\ket{10}$ are coupled through the Hamiltonian while the basis states $\ket{00}$ and $\ket{11}$ are decoupled from the quantum system. If we further require $\int^{\tau}_{0}J(t)\dd{t}=\int^{4\tau}_{3\tau}J(t)\dd{t}=\pi/4$ and $\int^{3\tau}_{\tau}J(t)\dd{t}=\pi/2$, the time evolution reads
	\begin{align}
		\text{$\mathcal{U}(t)=$} \left\{
		\begin{array}{l}
			e^{-i\int_{0}^{t}H^{\mathrm{eff}}_{\mathrm{S}}(t^{\prime})\dd{t^{\prime}}}, \quad \quad \quad ~~~~~t\in[0,\tau), \\
			e^{-i\int_{0}^{t}H^{\mathrm{eff}}_{\mathrm{S}}(t^{\prime})\dd{t^{\prime}}}\mathcal{U}(\tau), ~~~~~\quad  t\in\left[\tau,3\tau\right],
			\\
			e^{-i\int_{0}^{t}H^{\mathrm{eff}}_{\mathrm{S}}(t^{\prime})\dd{t^{\prime}}}\mathcal{U}(3\tau), ~~~\quad  t\in\left(3\tau,4\tau\right].
		\end{array}
		\right.
	\end{align}
	Governed by the piecewise Hamiltonian $H^{\mathrm{eff}}_{\mathrm{S}}(t)$ along with the requirement for the evolution time in each interval, the  quantum state in one of the states $\ket{+}=(\ket{01}+i\ket{10})/\sqrt{2}$ and $\ket{-}=(\ket{01}-i\ket{10})/\sqrt{2}$ will evolve according to
	\begin{align}
		&\ket{+}\xrightarrow{\mathcal{U}(\tau)}
		\ket{01}\xrightarrow{\mathcal{U}(\tau,3\tau)} e^{i(\pi/2-\gamma)}\ket{10}\xrightarrow{\mathcal{U}(3\tau,4\tau)}e^{-i\gamma}\ket{+}, \notag\\
		&\ket{-}\xrightarrow{\mathcal{U}(\tau)}
		-i\ket{10}\xrightarrow{\mathcal{U}(\tau,3\tau)}e^{i\gamma}\ket{01}\xrightarrow{\mathcal{U}(3\tau,4\tau)}e^{i\gamma}\ket{-},
	\end{align}
	and finally it will acquire a phase $-\gamma$ or $\gamma$.
	In this process, it is easy to verify that 
	$\bra{\psi(t)}H^{\mathrm{eff}}_{\mathrm{S}}(t)\ket{\psi(t)}=0$, where $\ket{\psi(0)}\in\{\ket{+},\ket{-}\}$. It implies that the phases $\gamma$ and $-\gamma$ are purely geometric phases.
	Meanwhile, the quantum state residing in the subspace spanned by $\{\ket{00},\ket{11}\}$ remains unchanged during the whole evolution.
	As a consequence, the unitary operator with the above design yields a geometric gate.  In the computational basis $\{\ket{00},\ket{01},\ket{10},\ket{11}\}$, the geometric gate can be expressed  as 
	\begin{align}
		\mathcal{U}(4\tau)=\left(
		\begin{array}{cccc}
			1   &0    &0   &0 \\
			0 &\cos\gamma &-\sin\gamma &0 \\
			0 &\sin\gamma &\cos\gamma &0 \\
			0   &0    &0   &1 
		\end{array}
		\right).
	\end{align}
	Obviously, it is an entangling two-qubit gate that we aim to realize.
	\begin{figure}[t]
		\includegraphics[scale=0.58]{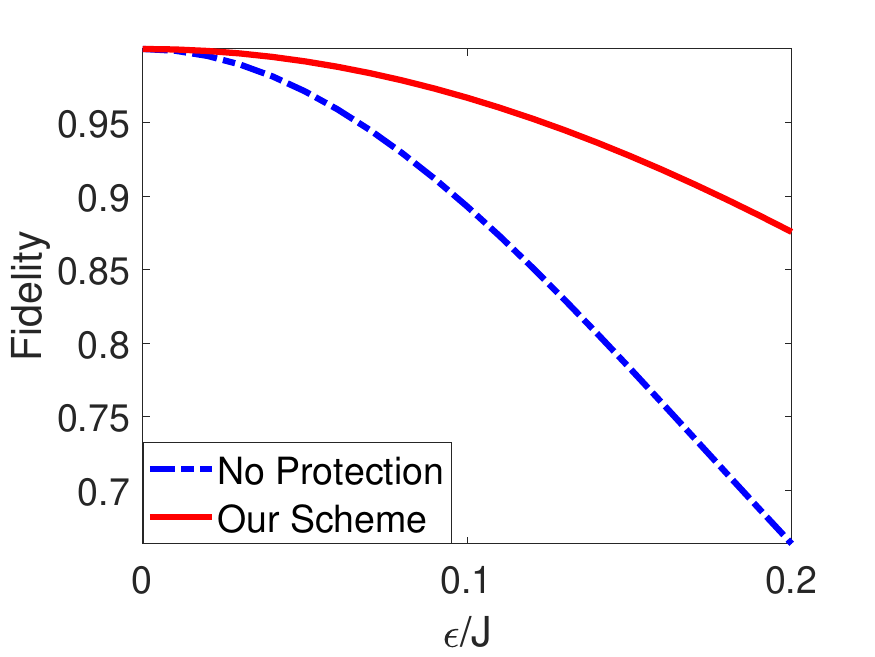}
		\caption{The fidelity of the two-qubit gate versus the noise strength $\epsilon$ over $[0,0.2J]$ under our scheme (red line) and the one without any protection (blue line).}
		\label{Fig2}
	\end{figure} 
	
	To evaluate the performance of our scheme, we also numerically calculate the fidelity of the quantum gate using our approach and the one without any protection as a reference. Similar to the one-qubit case, the system-environment Hamiltonian is still set as the Heisenberg coupling with the coupling strength $\epsilon$. In our simulations, we take the phase $\gamma=\pi/4$ and the qubit-qubit coupling strength $J=0.05\times2\pi\ \mathrm{MHz}$. Then, the driving Hamiltonian can be obtained by following the aforementioned method and the quantum gate yields	
		$\ket{00}\bra{00}+\exp(-i\pi/4)\ket{+}\bra{+}+\exp(i\gamma)\ket{-}\bra{-}+\ket{11}\bra{11}$. Additionally, we take the initial state as $\ket{10}$ and the strength of the square-shaped decoupling pulse as $2\pi\ \mathrm{MHz}$. In Fig.~\ref{Fig2}, we plot the fidelity $F$ versus the noise strength $\epsilon\in[0,0.2J]$ under decoupling protection and the one without any  protection, depicted by the red and blue lines. The result clearly shows that our scheme indeed enhances the performance of gate operations as our expectation.
	
	Until now, we have realized a universal set of geometric gates protected by dynamical decoupling, including arbitrary one-qubit gates and an entangling two-qubit gate. This completes our demonstration. 
	
	\section{Conclusion}
	
	In conclusion, we have proposed a scheme for the realization of geometric quantum computation protected by dynamical decoupling.
	Different from the previous scheme based on logical qubits, our scheme is implemented by using only physical qubits rather than logical qubits. This undoubtedly avoids the additional consumption of physical-qubit resources. Considering that the manipulation of physical qubits is much easier than that of logical qubits, our scheme naturally relaxes the restrictions imposed on the driving Hamiltonian for satisfying both the commutation relation and the geometric evolution requirement compared with the previous schemes. Moreover, our scheme focuses on the most general interaction between an individual qubit and its environment so that it mitigates not just dephasing noise but rather regular decoherence. All these merits indicate that our proposal represents a more realistic and effective approach towards the realization of geometric control with decoherence mitigation. 
	
	\begin{acknowledgments}	
		This work is supported by the National Natural Science Foundation of China through Grant No. 12305021.
	\end{acknowledgments}

\end{document}